\documentclass[nofootinbib,notitlepage,superscriptaddress,11pt,aps,prd,onecolumn]{revtex4-1}

\def\be{\begin{equation}}
\def\ee{\end{equation}}
\def\bea{\begin{eqnarray}}
\def\eea{\end{eqnarray}}

\usepackage{tensor}
\usepackage{amsmath,mathtools}
\usepackage{amssymb}
\usepackage{float}
\newcommand{\leri}[1]{\left(#1\right)}

\begin{document}
\vspace*{4cm}
\title{Quasinormal modes of Schwarzschild black holes in metric affine Chern-Simons theory}

\author{ F. BOMBACIGNO }

\address{Departament de F\'{i}sica Teòrica and IFIC\\ Centro Mixto Universitat de València- CSIC, Universitat de València
\\Burjassot 46100, València, Spain}
\begin{abstract}
We reformulate the Chern-Simons {modified gravity} in the metric-affine formalism, by enlarging the Pontryagin density with homothetic curvature terms which restore projective invariance without spoiling topologicity. The latter is then violated by promoting the coupling of the Chern-Simons term to a {(pseudo)}-scalar field. We derive the perturbative solutions for torsion and nonmetricity from the background fields, and we describe the dynamics for the resulting linearized metric and the scalar fields in a Schwarzschild black hole background. Then, by adopting numerical techniques we compute the quasinormal mode spectrum and the late-time tails for scalar and metric perturbations.
\end{abstract}
\maketitle

\section{Introduction}
Chern-Simons modified gravity (CSMG) was formulated by Jackiw and Pi \cite{Jackiw:2003pm}, in analogy with Chern-Simons modification of electrodynamics \cite{Carroll:1989vb}, and amounts to adding to the General Relativity Lagrangian the gravitational Pontryagin density, defined by $\,^{*}RR\equiv\,^{*}R^{\mu\nu\alpha\beta}R_{\nu\mu\alpha\beta}$, coupled to a pseudo-scalar field $\theta(x)$. The relevance of CSMG relies on the possibility of theoretically justify deviations from Kerr metric as proposed in \cite{Johannsen:2010xs}, by virtue of the odd parity property of the Pontryagin density, which makes it relevant in axially symmetric (rotating) spacetimes. The study of CSMG is also motivated by the presence of a CS term in particle physics \cite{Alexander:2009tp}, where it is required for dealing with the gravitational anomaly, or in string theory where it appears via the Green-Schwarz mechanism in low energy effective string models \cite{Adak2012}. Interestingly, CS corrections arise within Loop Quantum Gravity as well, when addressing chiral anomaly of fermions and the Immirzi field ambiguity \cite{Mercuri:2009zi}. The theory, indeed, might help in conceiving strategies to {probe (local) Lorentz/CPT} symmetry breaking in gravitation, and CS parity violation effects (birefringence, CMB polarization, baryon asymmetry problem) are already known in the literature \cite{Nojiri:2019nar,Bartolo:2017szm,PhysRevLett.96.081301}. Here we present a metric-affine generalization of standard CSMG, i.e. we assume the metric and the connection as independent variables, by requiring the projective invariance and the topologicity for the modified Pontryagin density. For all details, we remind the reader to the original paper \cite{Boudet:2022wmb}.

\section{The projective invariant CSMG model}
The starting point of our work is
\begin{equation}
S=\frac{1}{2\kappa^2}\int d^4x\sqrt{-g}\leri{\mathcal{R}+\frac{\alpha}{8}\theta(x)\varepsilon\indices{^{\mu\nu\rho\sigma}}\leri{\mathcal{R}\indices{^\alpha_{\beta\mu\nu}}\mathcal{R}\indices{^\beta_{\alpha\rho\sigma}}-\frac{1}{4}\hat{\mathcal{R}}\indices{_{\mu\nu}}\hat{\mathcal{R}}\indices{_{\rho\sigma}}} -\frac{\beta}{2}\nabla_\mu \theta \nabla^\mu\theta},
\label{action CS}
\end{equation}
where the Riemann tensor, the Ricci scalar and the homothetic curvature are given by
\begin{equation}
    \mathcal{R}\indices{^\rho_{\mu\sigma\nu}}=\partial_\sigma\Gamma\indices{^\rho_{\mu\nu}}-\partial_\nu\Gamma\indices{^\rho_{\mu\sigma}}+\Gamma\indices{^\rho_{\tau\sigma}}\Gamma\indices{^\tau_{\mu\nu}}-\Gamma\indices{^\rho_{\tau\nu}}\Gamma\indices{^\tau_{\mu\sigma}},\quad\mathcal{R}\equiv  g^{\mu\nu}\mathcal{R}\indices{^\rho_{\mu\rho\nu}},\quad \hat{\mathcal{R}}_{\mu\nu}\equiv\mathcal{R}\indices{^\rho_{\rho\mu\nu}}.
\end{equation}
Eq.~\ref{action CS} is invariant under the projective transformation
\begin{equation}
    \Gamma\indices{^\rho_{\mu\nu}}\rightarrow\tilde{\Gamma}\indices{^\rho_{\mu\nu}}=\Gamma\indices{^\rho_{\mu\nu}}+\delta\indices{^\rho_\mu}\xi_\nu,
    \label{projective}
\end{equation}
and we deal with torsion and nonmetricity, defined by:
\begin{equation}
    T\indices{^\rho_{\mu\nu}}\equiv\Gamma\indices{^\rho_{\mu\nu}}-\Gamma\indices{^\rho_{\nu\mu}},\qquad Q\indices{_{\rho\mu\nu}}\equiv-\nabla_\rho g_{\mu\nu}.
\end{equation}
An explicit solution for the connection can be obtained perturbatively, by expanding $\theta(x)=\bar{\theta}+\delta\theta(x)$ and $g_{\mu\nu}=\bar{g}_{\mu\nu}+h_{\mu\nu}$, where $\bar{\theta}$ is a constant and $\bar{g}_{\mu\nu}$ is a background metric which solves standard GR equations. Then, at the first order nonmetricity is vanishing and torsion results in
\begin{equation}
     \tensor[^{(1)}]{T}{_{\rho\mu\nu}}=\frac{\alpha}{2}\varepsilon\indices{^{\alpha\beta\gamma}_\rho} \bar{R}_{\mu\nu\beta\gamma}\nabla_\alpha\delta\theta.
     \label{solution first order torsion}
\end{equation}
By means of Eq.~\ref{solution first order torsion} we can evaluate the equations for the metric and scalar perturbations, which on the Schwarzschild background
\begin{equation}
    \bar{g}_{\mu\nu}dx^\mu dx^\nu = - \leri{1-\frac{2m}{r}} dt^2 + \leri{1-\frac{2m}{r}}^{-1} dr^2 + r^2 d\vartheta^2 + r^2 \sin^2\vartheta d\varphi^2,
\end{equation}
take the form, respectively
\begin{eqnarray}
    \frac{d^2Q}{dr_*^2} + \left[ \omega^2 -f\left(\frac{l(l+1)}{r^2} - \frac{6m}{r^3}\right)\right]Q & = & -\frac{6 \alpha m i \omega}{r^5} f \Theta,\label{axial metric mode}\\
    \leri{\beta + \frac{12\alpha^2m^2}{r^6}}\leri{\frac{d^2\Theta}{dr_*^2} + \left[ \omega^2 -f\left(\frac{l(l+1)}{r^2} + \frac{2m}{r^3}\right)\right]\Theta}+& &\nonumber\\-\frac{72\alpha^2 m^2 }{r^7}f \frac{d\Theta}{dr_*} + \frac{36\alpha^2 m^2}{r^8}f(2f-l(l+1))\Theta & = & \frac{6\alpha m }{-i\omega r^5}\frac{(l+2)!}{(l-2)!} f Q.\label{scalar mode}
\end{eqnarray}
where we adopted the Regge-Wheeler gauge \cite{Maggiore:2018sht} and tensor spherical harmonics decomposition, with $r_*= r + 2m\ln(r/2m - 1)$ the tortoise coordinate. Here $Q$ and $\Theta$ take care respectively of the axial part of metric and of the pseudo-scalar perturbation. We stress that while the metric perturbation equation retains the same form as in the metric case, Eq.~\ref{scalar mode} is instead endowed with additional terms, which still guarantee for $\beta=0$ a dynamics for $\Theta$ and also affect its asymptotic behavior. 
\newpage
\section{Numerical results for quasinormal mode frequencies}
\begin{figure}[t]
  \centering
  \includegraphics[width=0.8\linewidth]{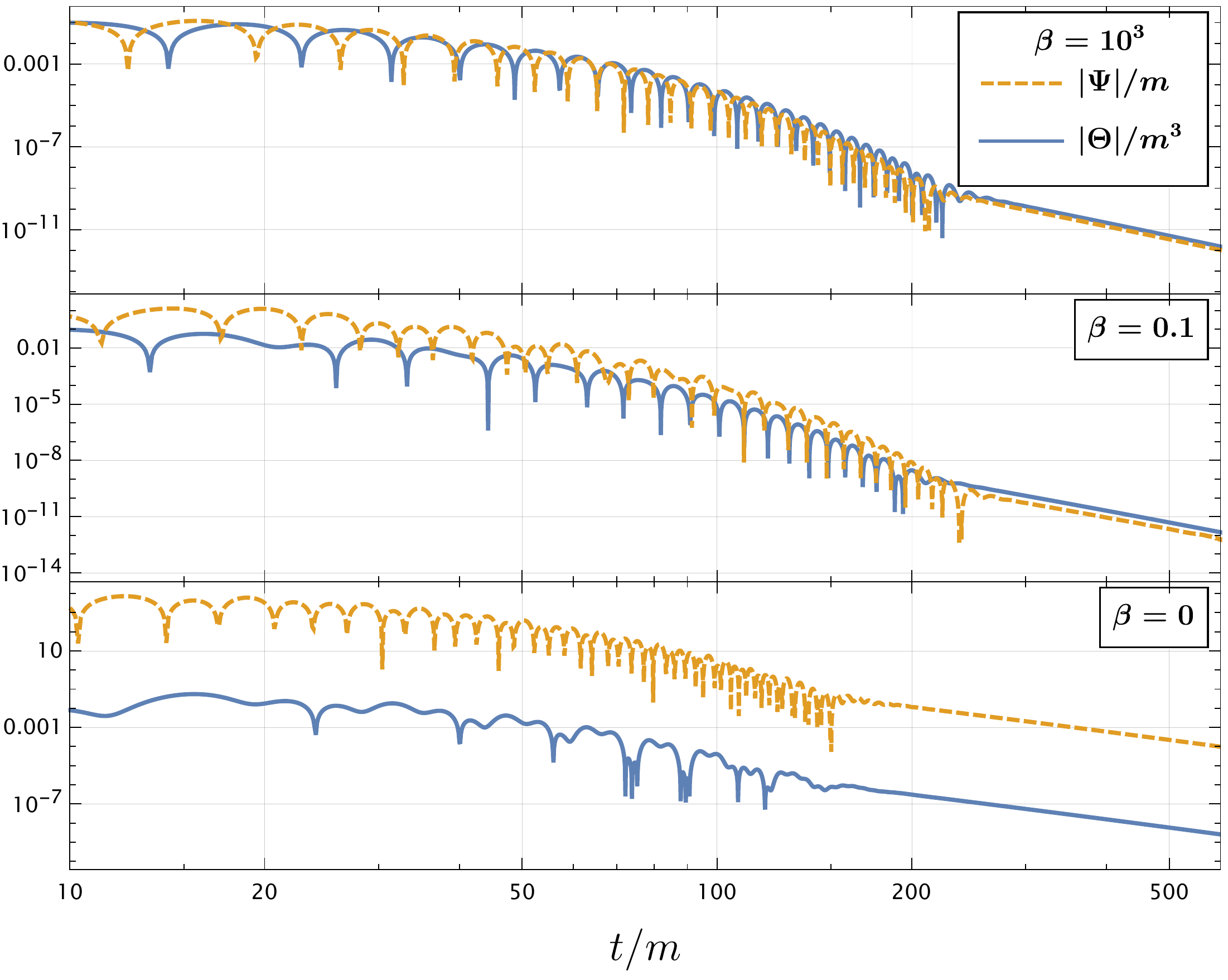}
\caption{Evolution of metric (continuous) and scalar (dashed) perturbations as a function of $t/m$ for $\beta=10^3$ (top), $\beta = 0.1$ (center) and $\beta = 0$ (bottom). Straight lines represent a power law behavior.}
\label{fig: QNMs}
\end{figure}

\begin{figure}[H]
  \centering
  \includegraphics[width=0.8\linewidth]{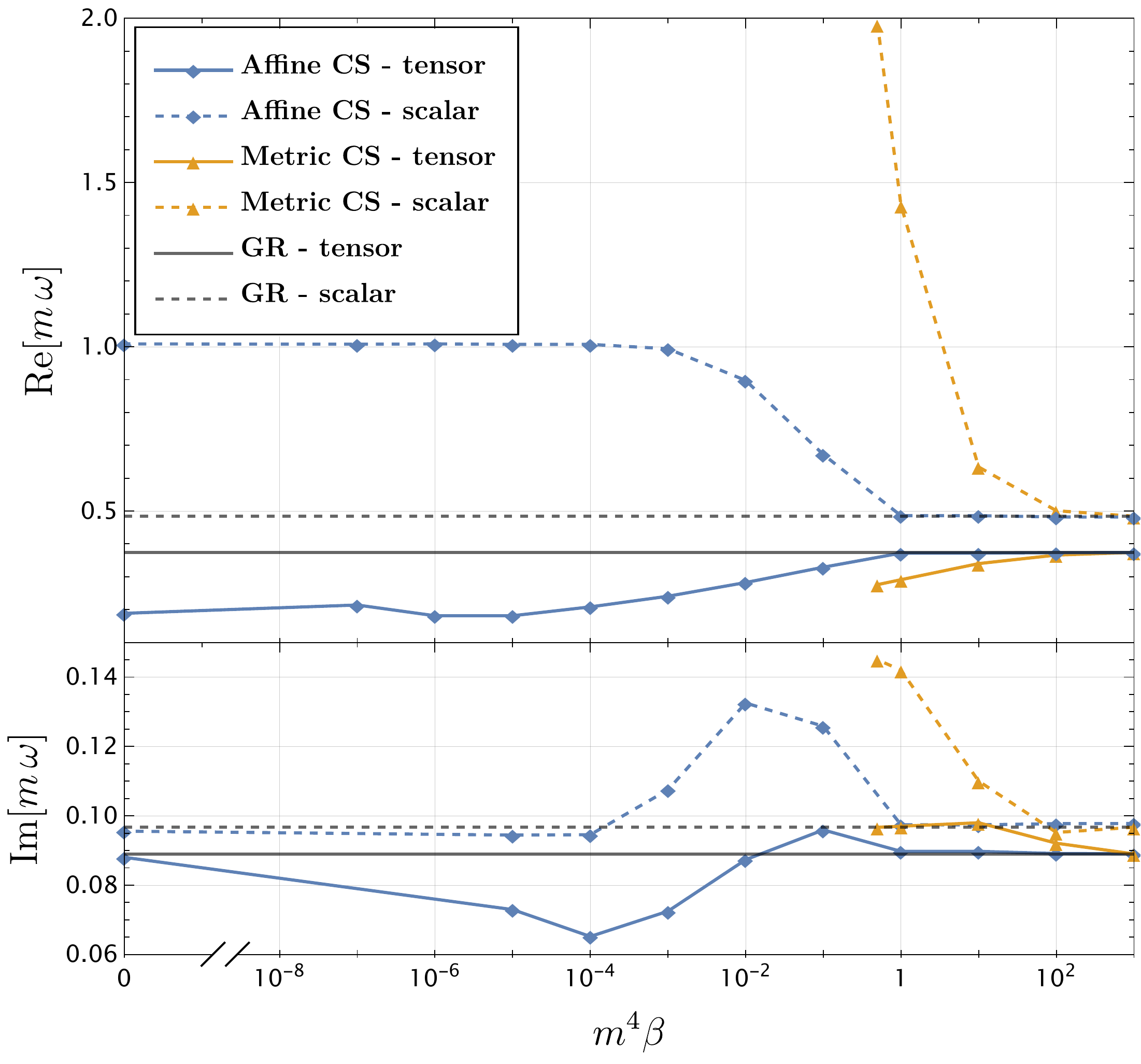}
\caption{Real (top) and imaginary (bottom) part of quasinormal frequencies of the fundamental $l=2$ tensor (continuous) and scalar (dashed) modes as a function of $\beta$, for metric (data taken from \cite{PhysRevD.81.124021}) and metric-affine CSMG.}
\label{fig: freq}
\end{figure}

When $\beta$ is very large, the scalar field equation decouples from the metric one and we obtain a Klein-Gordon equation on a Schwarzschild background. In this case the scalar perturbation is described by single mode oscillations with the same frequencies as in GR. Conversely, since the metric perturbation is always coupled to the scalar field, it is endowed with a superposition of the gravitational mode and the scalar one, which are given, respectively, by $\omega_g = 0.37 - i\, 0.089,\quad\omega_s = 0.48 - i\, 0.097$. Then, when $10^{-1}\lesssim\beta \lesssim 10$, the values of the two frequencies are nearly unchanged and now both perturbations oscillate with a superposition of the two modes. This two modes behavior continues to hold for lower values of $\beta$, even if the frequencies deviate from the GR values. In particular for $\beta$ down to the limit case $\beta=0$, they seem to saturate to stable values, as shown in Fig. \ref{fig: freq}. 
\\The late-time behavior of perturbations is described by power law tails $\sim  t^{-\mu}$, and for $\beta\neq 0$ results are indistinguishable from GR, where given the angular number $l$, the following holds $\mu=2l+3$. For $\beta=0$, instead, the relation between $\mu$ and $l$ is now consistent with $\mu = 0.884 \, l + 2.78$ (Fig.~\ref{fig: QNMs}).


\section{Discussion}
The lowest QNM ($l=2$) discloses a deviation from both GR and metric {CSMG}, according the value of the parameter $\beta$ (Fig.~\ref{fig: freq}). With respect to metric {CSMG}, the deviation begins at smaller values of $\beta$, implying that GR results are reproduced for a larger range of $\beta$ values, and the superposition of two modes is always present. However, if the two frequencies were detected, they would be seen as the $l=2$ tensor and scalar modes in {CSMG} or as the $l=2$ and $l=3$ tensor modes of GR \cite{PhysRevD.81.124021}. In this regard, the signal to noise ratio of the detector, required for separating the two frequencies for $\beta=0$, is lower with respect to GR and metric CSMG case, due to the larger gap between the frequencies \cite{PhysRevD.73.064030} (See Fig.~\ref{fig: freq}). Moreover, for $\beta=0$ also the late-time signal would represent a clear observational mark, by virtue of the different power tail exponent.
\section*{Acknowledgments}
The work of F. B. has been supported by the Fondazione Angelo della Riccia grant and by the project PROMETEO/2020/079 (Generalitat Valenciana).

\section*{References}
\bibliography{references}





\end{document}